\documentclass[pra]{revtex4-2}

\usepackage{graphicx}
\usepackage{dcolumn}
\usepackage{bm}

\begin{document}

\title{Energy-time entanglement and intermediate state dynamics in two photon absorption}

\author{Baihong Li}
 \altaffiliation[li-baihong@163.com]{}
\author{Holger F. Hofmann}%
 \email{hofmann@hiroshima-u.ac.jp}
\affiliation{%
Graduate School of Advanced Science and Engineering, Hiroshima University, Kagamiyama 1-3-1,
Higashi, Hiroshima 739-8530, Japan
}%

\begin{abstract}
It is well known that energy-time entanglement can enhance two photon absorption (TPA) by simultaneously optimizing the two photon resonance and the coincidence rate of photons at the absorber. However, the precise relation between entanglement and the TPA rate depends on the coherences of intermediate states involved in the transition, making it a rather challenging task to identify universal features of TPA processes. In the present paper, we show that the theory can be simplified greatly by separating the two photon resonance from the temporal dynamics of the intermediate levels. The result is a description of the role of entanglement in the TPA process by a one-dimensional coherence in the Hilbert space defined by the arrival time difference of the two photons. Transformation into the frequency difference basis results in Kramers-Kronig relations for the TPA process, separating off-resonant contributions of virtual levels from resonant contributions. In particular, it can be shown that off-resonant contributions are insensitive to the frequencies of the associated virtual states, indicating that virtual-state spectroscopy of levels above the final two photon excited state is not possible.
\end{abstract}

\maketitle

\section{Introduction}
\label{sec:intro}

It has been long known that the high coincidence rates of down-converted photon pairs are particularly well suited for two photon absorption (TPA), resulting not only in a linear dependence of TPA rate on photon flux \cite{Gea89,Jav90}, but also on additional enhancements of the TPA rate due to energy-time entanglement \cite{You09,Oka10,Oka18a,Oka18b}. The experimental verification of this features in atomic \cite{Geo95,Day04,Day05} and molecular \cite{Lee06,Guz10,Guz13} samples has established the use of entangled photons in non-linear quantum spectroscopies such as virtual-state spectroscopy \cite{Per98,Sal98,Leo13,Svo18,Svo19}, pump-probe spectroscopy \cite{Sch16a,Sch16b,Sch17a,Sch17b,Sch18} and two-dimensional spectroscopy\cite{Ric10,Ros13}.

The precise relation between the energy levels of a material and the quantum coherence of the two photon wave function of photons emitted by an energy-time entangled source in a TPA process can be rather complicated since it involves quantum coherences between all of the intermediate levels participating in the electronic response of the absorber. The initial analysis of this problem was based on the expectation that the comparatively simple spectral features of entangled light can be used to reveal details of the level structures of transient states excited during the absorption process \cite{Per98,Sal98,Leo13}. However, this approach has left many open questions regarding the precise role of the entanglement in the TPA process. In particular, it would seem that the possibility of TPA enhancement depends critically on the specific level structure of the two photon absorber \cite{Leo13}. It is therefore important to clarify the general relation between temporal two photon coherences in the absorbed light and the dynamics of TPA determined by the level structure of the absorbing material. 

In this paper, we show that the theory of TPA can be simplified greatly by separating the two photon resonance from the temporal dynamics of the intermediate levels. The result is a description of the role of entanglement in the TPA process by a one-dimensional coherence in the Hilbert space defined by the arrival time difference of the two photons. Our theory is based on the description of TPA as an effective projection of the two photon input states onto an optimally absorbed two photon state defined by the dynamics of the dipole response. If the level broadening of the final excited state of the TPA process can be neglected, it is possible to separate this optimally absorbed two photon state into a delta function of the sum frequency and a wave function of the arrival time difference defined by the dynamics of the intermediate levels. Since the medium always evolves in the forward direction in time, the wave function depends on the absolute value of the time difference. In the frequency difference representation, this results in off-resonant contributions to the TPA described by Kramers-Kronig relations. Based on this result, we can identify and eliminate unphysical negative frequency contributions to the optimally absorbed state. In particular, this elimination removes the resonant contributions of levels with energies higher than the maximal available single photon energy in the input state. Our analysis thus shows that the original proposal of virtual-state spectroscopy of levels above the final two photon excited state proposed by Saleh et al. \cite{Sal98} inadvertently included unphysical negative frequency components in the input state. Once these unphysical components are removed, the frequencies of levels above the two photon excited state do not show up in the dependence of TPA probabilities on time differences between the input photons. Instead we can show that the TPA probability of levels with energies higher than the final excited states contribute a time difference dependence given by a sinc function characterized only by the frequency of the two photon resonance. The optimally absorbed state is therefore an ideal broad band entangled state with a rectangular spectrum of frequency differences. 

The rest of the paper is organized as follows. In section \ref{sec:states}, we introduce the perturbation theory of TPA and show that the probability of TPA can be given by an effective projection onto an optimally absorbed two photon state defined by the dipole dynamics of the absorbing material. In section \ref{sec:dynamics}, we analyze the dynamics of the intermediate levels and show that the optimally absorbed two photon state can be expressed by an energy-time entangled state with a well-defined sum frequency and a one-dimensional wave function in the Hilbert space of arrival time difference. The dependence of the amplitude of the wave function on arrival time difference is given by the time evolution of an effective intermediate state generated by the initial absorption event and its overlap with the effective final state formed by the coherent superposition of the transition matrix elements of the dipole operator. In section \ref{sec:KKR}, we determine the frequency difference representation of the optimally absorbed two photon state and identify resonant and off-resonant contributions based on the Kramers-Kronig relations representing the causality of the time evolution. In section \ref{sec:negatives}, we remove the unphysical negative frequencies in the frequency representation of the optimally absorbed state and discuss the impossibility of observing resonances for energy levels higher than the final two photon excited state. Section \ref{sec:conclude} summarizes the results and concludes the paper.

\section{Temporal coherence in two photon absorption}
\label{sec:states}

The basic theory of TPA is well established and has been studied in a wide range of contexts \cite{Per98,Leo13,Sch17a}. However, most of these studies have focused on the variety of possible level structures in the material while describing the two photon coherences of the input in terms of a small selection of rather simple two photon states. It has therefore been difficult to identify the precise relation between two photon coherences in the input state and the dynamics of the multi-level system describing a two photon absorber. In the present paper, we set out to formulate the problem by presenting a general framework for the analysis of quantum coherence in TPA that is still simple enough to identify the general features shared by TPA processes using different materials and different light sources. 

In order to analyze the temporal coherence in TPA, it is convenient to present the theory of TPA in the time domain. In a TPA process, the material system starts out in its ground state $\mid g \rangle$ and is excited to its final state $\mid f \rangle$ by a two step process involving intermediate levels $\mid m \rangle$. If the material system interacts with sufficiently weak fields, second-order time-dependent perturbation theory can be used \cite{Per98,Sal98,Leo13} and the probability of TPA can be expressed by
\begin{equation}
\label{eq:TPA}
P_{\mathrm{TPA}} = \left|{\frac{1}{\hbar^2}\ \int \int\,M_{\mathrm{mat.}}(t_1,t_2)M_{\mathrm{field}}(t_1,t_2) dt_2\,dt_1}\right|^2 \, .
\end{equation}
Here, the quantum coherence of the input field is characterized by the two time correlation function of the field, $M_{\mathrm{field}}$, and the coherent response of the material is characterized by the two time correlation function of the dipole response, $M_{\mathrm{mat.}}$. Effectively, perturbation theory expresses TPA by a linear response function characterized by a two dimensional time dependence. 

We first consider the characteristics of the optical input which are represented by the two time correlation function $M_{\mathrm{field}}$ of the input light field. In general, this two time correlation can be determined from an arbitrary quantum state of the input light field by applying the local field operators at times $t_1$ and $t_2$. For a two photon state $\mid \psi_{TP} \rangle$, the application of field operators corresponds to the annihilation of one photon each, so that the output state of TPA is the vacuum state. The electric field amplitudes associated with a local photon absorption event at a time $t$ depend on the mode volume defined by the beam profile \cite{You09}. 
Since the input field is not confined to a resonator, it is necessary to convert the power $P_0$ of the input beam into a photon current $\gamma_{\mbox{flux}}$. This conversion can be achieved by dividing $P_0$ by the average photon energy $\hbar \omega_{\mathrm{av.}}$. According to classical electromagnetism, the expectation value of the squared electric field  at the center of the beam $\langle E^2_{\mathrm{loc.}} \rangle$ is proportional to the power $P_0$ as described by the surface integral of the Poynting vector for the specified beam profile. The relation between the field strength at the position of the absorber and the photon current in the beam can then be described by the ratio $J_Q=\langle E^2_{\mathrm{loc.}} \rangle/\gamma_{\mbox{flux}}$. This ratio converts the temporal wave function of photons into an electric field amplitude. When applied to the two photon wave function $\langle t_1, t_2 \mid \psi_{TP} \rangle$ in the time basis $\mid t_1, t_2 \rangle$ the ratio $J_Q$ converts the coherences between the photon arrival times $t_1$ and $t_2$ into the two time correlation function of the input field,  
\begin{equation}
\label{eq:flux}
M_{\mathrm{field}}(t_1,t_2) = J_Q \langle t_1,t_2 \mid \psi_{TP} \rangle.
\end{equation}
The beam profile constant $J_Q$ thus describes the relation between photon wave functions and the electric field amplitudes responsible for the dipole transitions in the absorbing material.

The response of the material is determined by the dynamics of electric dipoles describing the transitions between the energy levels of the absorber. Since the two time correlation of the field is proportional to the two photon wave function as shown in Eq.(\ref{eq:flux}), the two time correlation of the material dipole defines an optimal two photon coherence given by a two photon state $\mid \mu_{\mathrm{abs.}}(0) \rangle$. The TPA probability can than be expressed as a projection of the input state $\mid \psi_{TP} \rangle$ onto the optimally absorbed two photon state,
\begin{equation}
P_{\mathrm{TPA}} = \left| \int \int\, \langle \mu_{\mathrm{abs.}}(0) \mid t_1, t_2 \rangle \langle t_1, t_2 \mid \psi_{TP} \rangle dt_2\,dt_1\right|^2 \, ,
\end{equation}
where comparison with Eq.(\ref{eq:TPA}) shows that the optimally absorbed state $\mid \mu_{\mathrm{abs.}}(0) \rangle$ is determined by the two time correlations of the dipoles in the material,
\begin{equation}
\label{eq:optimum}
\langle \mu_{\mathrm{abs.}}(0) \mid t_1, t_2 \rangle = \frac{J_Q}{\hbar^2} M_{\mathrm{mat.}}(t_1,t_2).
\end{equation}
It should be noted that the two photon state $\mid \mu_{\mathrm{abs.}}(0) \rangle$ is not normalized. In fact, the maximal TPA probability should be much lower than one to ensure that the perturbation analysis of Eq.(\ref{eq:TPA}) is valid and higher order processes can be neglected. For realistic situations, we can therefore assume that
\begin{equation}
\langle \mu_{\mathrm{abs.}}(0)\mid \mu_{\mathrm{abs.}}(0) \rangle \ll 1.
\end{equation}
As suggested by Eq.(\ref{eq:optimum}), the precise normalization depends on the beam profile constant $J_Q$, indicating the dependence on TPA rates on the spot size of the beam at the absorber.

The dipole dynamics of TPA defines an optimal temporal coherence of the two photon wave function that achieves the maximal possible absorption probability for a given beam profile. If the maximal absorption cross-section is known, the TPA rates can serve as a measure of fidelity for the two photon state $\mid \mu_{\mathrm{abs.}}(0) \rangle$ in the two photon wave unction of the input photons. In general, TPA rates increase with the quantum state fidelity of $\mid \mu_{\mathrm{abs.}}(0) \rangle$ in the input two photon wave function.

\section{Energy-time entanglement and intermediate state dynamics}
\label{sec:dynamics}

The coherent response of the material can be described by the dipole correlation function $M_{\mathrm{mat.}}(t_1,t_2)$. Assuming that the TPA happens on time scales much shorter than the dephasing times of the intermediate levels $\mid m \rangle$, the dipole correlation can be expressed using the dipole operator $\hat{d}$ describing transitions between the ground state $\mid g \rangle$ and the intermediate level $\mid m \rangle$, and between the intermediate level $\mid m \rangle$ and the final state $\mid f \rangle $. 
The dipole correlation function depends on the time evolution of the intermediate levels $\mid m \rangle$ given by their energies $\hbar \omega_m$, where the initial time $\mbox{Min}(t_1,t_2)$ describes the transition from $\mid g \rangle$ to $\mid m \rangle$ and the final time $\mbox{Max}(t_1,t_2)$ describes the transition from $\mid m \rangle$ to $\mid f \rangle$. The dipole correlation function can thus be expressed in terms of the dynamics of the levels involved in the TPA process,
\begin{equation}
M_{\mathrm{mat.}}(t_1,t_2) = \sum_m \langle f \mid \hat{d} \mid m \rangle \langle m \mid \hat{d} \mid g \rangle \; \exp(-i \omega_g \mbox{Min}(t_1,t_2) - i \omega_m |t_2-t_1| + i \omega_f \mbox{Max}(t_1,t_2)),
\end{equation}
where the energies of the ground state $\mid g \rangle$ and the final state $\mid f \rangle$ are given by $\omega_g$ and $\omega_f$ respectively, and the energies of the intermediate levels are given by $\omega_m$. 
A more consistent mathematical description is obtained by expressing the minimal and the maximal values of $t_1$ and $t_2$ using the average time $(t_1+t_2)/2$ and the absolute value of the time difference $|t_2-t_1|$. The dipole correlation function then reads
\begin{equation}
\label{eq:resp}
M_{\mathrm{mat.}}(t_1,t_2) = \sum_m \langle f \mid \hat{d} \mid m \rangle \langle m \mid \hat{d} \mid g \rangle \;\exp(i \omega_{\mathrm{av.}} |t_2-t_1|) \exp(i \omega_{gf} \frac{t_1+t_2}{2}),
\end{equation}
where $\omega_{\mathrm{av.}}=(\omega_f+\omega_g)/2$ is the average frequency of the ground state and the excited state and $\omega_{gf}=\omega_f-\omega_g$ is the two photon transition frequency.

It is possible to understand the dynamics described by the intermediate levels $\mu=\{ m \}$ by defining an initial quantum state $\mid \Phi_{g\mu} \rangle$ corresponding to the superposition of intermediate levels immediately after the absorption of the first photon and a final state $\mid \Phi_{\mu f} \rangle$ corresponding to the superposition of intermediate levels that achieves the largest transition dipole moment with regards to the final transition to the state $\mid f \rangle$. The definitions of the normalized states and their transition dipole moments are given by 
\begin{eqnarray}
d_{g \mu} \mid \Phi_{g \mu} \rangle &=& \sum_m \langle m \mid \hat{d} \mid g \rangle \mid m \rangle,
\nonumber \\
d^*_{\mu f} \mid \Phi_{\mu f} \rangle &=& \sum_m \langle m \mid \hat{d} \mid f \rangle \mid m \rangle.
\end{eqnarray}
The time evolution of the intermediate levels can then be expressed by the unitary operator
\begin{equation}
\label{eq:unitary}
\hat{U}(t) = \sum_m \exp(-i \omega_m t) \mid m \rangle\langle m \mid.
\end{equation}
The sum over the intermediate states $\mid m \rangle$ in Eq. (\ref{eq:resp}) can be summarized by a single time dependent transformation of the effective initial state $\mid \Phi_{g\mu} \rangle$. The dipole correlation is then given by the time dependence of the overlap between the initial and the final wave function of the material system,
\begin{equation}
M_{\mathrm{mat.}}(t_1,t_2) = d_{g \mu} d_{\mu f} \langle \Phi_{\mu f} \mid \hat{U}(|t_2-t_1|) \mid \Phi_{g \mu} \rangle \exp(i \omega_{\mathrm{av.}} |t_2-t_1|) \exp(i \omega_{gf}\frac{t_1+t_2}{2}).
\end{equation}
It is natural that the TPA process does not depend on the absolute photon arrival times but only on the time difference between the two photons, $|t_2-t_1|$. The average arrival time only appears in the phase of the dipole correlations, which establishes the frequency $\omega_{gf}$ as the resonant frequency of the two photon transition. It is convenient to transform the two photon time coordinates into collective coordinates representing the average time and the time difference,
\begin{eqnarray}
t_+ &=& \frac{1}{2}(t_1+t_2)
\nonumber \\ 
t_- &=& (t_2-t_1).
\end{eqnarray}
In this representation, the dipole correlation function reads
\begin{eqnarray}
\label{eq:mat}
M_{\mathrm{mat.}}(t_+,t_-) = &&d_{g \mu} d_{\mu f} \langle \Phi_{\mu f} \mid \hat{U}(|t_-|) \mid \Phi_{g \mu} \rangle 
\exp(i \omega_{\mathrm{av.}} |t_-|) \exp(i \omega_{gf} t_+),
\end{eqnarray}
In these coordinates, the dipole correlation function can be factorized into a harmonic oscillation at frequency $\omega_{gf}$ in the average time coordinate and a more complicated function of the time difference given by the unitary dynamics of the material system. It should be noted that the unitary dynamics is only sensitive to the absolute value of the time difference. This means that it does not matter which photons is absorbed first in the TPA process. The appearance of the absolute value $|t_-|$ in the material response is an important signature which determines the temporal coherence of the optimally absorbed two photon state. 

The optimal absorbed two photon state $\mid \mu_{\mathrm{abs.}}(0) \rangle$ can be obtained from the dipole correlation function according to Eq.(\ref{eq:optimum}). In the basis given by the average arrival time $t_+$ and the time difference $t_-$ between photon arrival times, the state is given by 
\begin{equation}
\label{eq:timestate}
\langle \mu_{\mathrm{abs.}}(0) \mid t_+, t_- \rangle = \sigma_{TP} \langle \Phi_{\mu f} \mid \hat{U}(|t_-|) \mid \Phi_{g \mu} \rangle \exp(i \omega_{\mathrm{av.}} |t_-|) \exp(i \omega_{gf} t_+).
\end{equation}
Here the absorption cross section $\sigma_{TP}$ describes the probabilities of transitions associated with the beam profile and the optimal intermediate states $\mid \Phi_{g \mu} \rangle$ and $\mid \Phi_{\mu f} \rangle$,
\begin{equation}
\sigma_{TP} = \frac{J_Q d_{g \mu} d_{\mu f}}{\hbar^2}.
\end{equation}
If perturbation theory is valid, the cross section $\sigma_{TP}$ ensures that the TPA probabilities are well below one, even if the two photon coherences are matched with the optimally absorbed state $\mid \mu_{\mathrm{abs.}}(0) \rangle$.

The wave function of the optimally absorbed state $\mid \mu_{\mathrm{abs.}}(0) \rangle$ can be separated into a product of a wave function of the average arrival time $t_+$ and a wave function of the time difference $t_-$. It is easy to see that this wave function describes entanglement between the two photons arriving at $t_1$ and at $t_2$. To be separable, the wave function needs to be given by a product of a function of $t_1$ and a function of $t_2$. Since the wave function in $t_+$ is given by a constant oscillation at frequency $\omega_{gf}$, the only possible function of the absolute value of arrival time difference $|t_-|$ that would permit a factorization into functions of $t_1$ and $t_2$ would be a constant function independent of $|t_-|$. Incidentally, this condition can be satisfied when there is a single level at $\omega_m=\omega_{\mathrm{av.}}$. In this case, the optimal TPA is obtained when both photons are resonant with the single transition frequency of $\omega_{\mathrm{av.}}$. For all other spectra of $\omega_m$, any photon pairs described by the optimal absorption state $\mid \mu_{\mathrm{abs.}}(0) \rangle$ will be entangled in energy and time.

It should be noted that the comparison between the absorption cross-sections of entangled photon pairs and photons from coherent sight sources is usually complicated by the technical difficulty of matching the sum frequency to a particular two photon resonance. This seems to be the main reason why previous theories did not consider the separation of sum frequency and frequency differences used here \cite{Ray21,Par21}. Even though the separation applied here might simplify some of the arguments in these previous works, its main merit is the identification of the role played by the intermediate levels $\mid m \rangle$ in the TPA process.

The time basis representation of the optimally absorbed two photon state $\mid \mu_{\mathrm{abs.}}(0) \rangle$ can be used to explain the effects of small time delays $\tau$ imposed on one of the two input photons on the probability of TPA. In the ideal limit of energy-time entanglement, the two photons are emitted at the same time and a time delay between then controls the time between the initial absorption and the final absorption in the material. Previous studies of this method of non-classical spectroscopy have suggested that this approach can be used to identify the complete spectrum of the intermediate levels involved in the TPA \cite{Sal98,Leo13}. It may therefore be useful to describe the effects of a time delay $\tau$ on the TPA probability as a modification of the optimally absorbed state defined by the material. It is in fact quite straightforward to introduce a time delay $\tau$ applied to input photon 2 in the theoretical description developed above. The time delay changes the actual arrival time of the second photon from $t_2$ to $t_2+\tau$ and the arrival time difference between the two photons from $t_-$ to $t_-+\tau$. With respect to these new time coordinates of the input photons, the optimally absorbed two photon state at a time delay of $\tau$ is given by
\begin{equation}
\langle \mu_{\mathrm{abs.}}(\tau) \mid t_+,t_- \rangle = \langle \mu_{\mathrm{abs.}}(0) \mid t_++\tau/2,t_-+\tau \rangle.
\end{equation}
It should be noted that the application of a time delay to only one of the photons implies that the two photons are distinguishable because they propagate along different optical paths. The wave function can still be separated into a wave function sensitive to the time difference and a continuous phase oscillation at frequency $\omega_{gf}$ for the average time, but the discontinuity of the time dependence now occurs at $t_-=-\tau$ instead of $t_-=0$,  
\begin{eqnarray}
\label{eq:delay}
\langle \mu_{\mathrm{abs.}}(\tau) \mid t_+, t_- \rangle =&& \sigma_{TP} \langle \Phi_{\mu f} \mid \hat{U}(|t_-+\tau|) \mid \Phi_{g \mu} \rangle \exp(i \omega_{\mathrm{av.}} |t_-+\tau|) \exp(i \omega_{gf} (t_++\tau/2)).
\end{eqnarray}
Independent of the time shift $\tau$, this state is an eigenstate of the sum frequency $\omega_+=\omega_{gf}$, where $\omega_+=\omega_1+\omega_2$ is the sum of the single photon frequencies $\omega_1$ and $\omega_2$. We can now consider an idealized input state $\mid E_{\mathrm{ideal}}\rangle$ that is a simultaneous eigenstate of a sum frequency of $\omega_+$ and a time difference of $t_-=0$. For this input state,
\begin{eqnarray}
\label{eq:ideal}
\langle \mu_{\mathrm{abs.}}(\tau) \mid E_{\mathrm{ideal}} \rangle =&& \sqrt{2 \pi} \sigma_{TP} \langle \Phi_{\mu f} \mid \hat{U}(|\tau|) \mid \Phi_{g \mu} \rangle \exp(i \omega_{\mathrm{av.}} |\tau|) \delta(\omega_+-\omega_{gf}).
\end{eqnarray}
Since phase factors do not show up in the probability $P_{\mathrm{TPA}}(\tau)$, the experimental evidence obtained with an ideal energy-time entangled state will correspond to the squared overlap between the internal state $\hat{U}(|\tau|) \mid \Phi_{g\mu} \rangle$ at a time $\tau$ after the first excitation and the final state $\mid \Phi_{\mu f} \rangle$ which represents the absorption of the second photon as an effective measurement of the intermediate state. The information most naturally obtained by applying a time delay to TPA of energy-time entangled photons is therefore equivalent to a projection measurement of the time evolution of the initial state  $\mid \Phi_{g \mu} \rangle$ onto a fixed target state $\mid \Phi_{\mu f} \rangle$ determined by the dipole transitions to the two photon excited state $\mid f \rangle$.

It may be useful to illustrate the role of time delays $\tau$ using the particularly simple case of two intermediate levels described by an average energy of $\hbar \omega_\mu$ and a splitting of $\hbar \Omega$, so that the two intermediate energy levels of the absorber have energies of $\hbar (\omega_\mu + \Omega/2)$ and $\hbar (\omega_\mu - \Omega/2)$, respectively. Such a scenario could be realized by using the Zeeman splitting of intermediate atomic levels. For equal dipole transition elements between $\mid g \rangle$ and the two intermediate levels and equal dipole transition elements between the two intermediate levels and $\mid f \rangle$ both $\mid \Phi_{g \mu} \rangle$ and $\mid \Phi_{\mu f} \rangle$ are given by equal superpositions of the two levels.  The explicit dependence of the optimally absorbed state in Eq.(\ref{eq:timestate}) on the arrival time difference $|t_-|$ is then given by 
\begin{equation}
\langle \mu_{\mathrm{abs.}}(0) \mid t_+, t_- \rangle = \sigma_{TP} \cos(\Omega |t_-|/2)
\exp(-i (\omega_\mu-\omega_{\mathrm{av.}}) |t_-|) \exp(i \omega_{gf} t_+).
\end{equation}
The interferences between the two intermediate states shows up as modulations of the amplitudes associated with the photon arrival time difference $t_-$. These modulations can be observed directly in the dependence of two photon absorption probability $P_{\mathrm{TPA}}$ on the time delay $\tau$ between two photons from the broadband entangled source described by $\mid E_{\mathrm{ideal}} \rangle$ in Eq.(\ref{eq:ideal}),
\begin{equation}
P_{\mathrm{TPA}} (\tau) = \frac{1}{2}\sigma_{TP}^2 \left(1 + \cos(\Omega \tau)\right).
\end{equation}
This example shows how the time delay between two broadband entangled photons can sample the quantum dynamics of intermediate levels. If more realistic input states are used, it is also possible to observe phase oscillations of $\langle \mu_{\mathrm{abs.}}(0) \mid t_+, t_- \rangle$ associated with individual intermediate energies $\hbar \omega_m$. 

The time representation summarizes the contributions of all intermediate levels in terms of the time dependence of a single intermediate state. In order to understand how each of the intermediate levels contribute to this dependence of TPA on time delays between the photons, it is useful to convert the results given above into the frequency basis. As will be shown later, this is particularly important for a realistic assessment of the contributions from levels above the two photon excited state $\mid f \rangle$.  

\section{Kramers-Kronig relations for two photon absorption}
\label{sec:KKR}

In principle, the transformation of the time basis into the frequency basis is a straightforward matter. Using the collective time coordinates $t_+$ and $t_-$, the corresponding collective frequencies are given by 
\begin{eqnarray}
\omega_+ &=& \omega_1+\omega_2,
\nonumber \\ 
\omega_- &=& \frac{1}{2}(\omega_2-\omega_1).
\end{eqnarray}
As we already remarked in the previous section, the optimal absorption state is an eigenstate of $\omega_+$ with an eigenvalue of $\omega_{gf}$. It is therefore convenient to define a one dimensional wave function $\Gamma(\omega_-)$ by separating the delta function representing the eigenstate of $\omega_+$,
\begin{equation}
\langle \mu_{\mathrm{abs.}}(0) \mid \omega_+, \omega_- \rangle = \Gamma(\omega_-) \delta(\omega_+ - \omega_{gf}).
\end{equation}
The delta function indicates that the optimally absorbed state should be resonant with the final state of the two photon transition. In reality, this resonance will be limited by the line width of the two photon excited state $\mid f \rangle$, but we will assume that this line width is sufficiently narrow to justify the approximate representation by a delta function. 

After the separation, the Fourier integral that determines $\Gamma(\omega_-)$ is given by 
\begin{equation}
\label{eq:FT}
\Gamma(\omega_-) = \sigma_{TP} \int \langle \Phi_{\mu f} \mid \hat{U}(|t_-|) \mid \Phi_{g \mu} \rangle \exp(i \omega_{\mathrm{av.}} |t_-|) \exp(i \omega_- t_-) \, dt_-.
\end{equation}
We can now identify the contributions of the different intermediate levels $\mid m \rangle$ using Eq.(\ref{eq:unitary}). The result is a sum over Fourier transforms determined by the energies $\nu_m$ that determine the resonances of these levels,
\begin{equation}
\label{eq:Gamma}
\Gamma(\omega_-) = \sum_m C_m \int \exp(-i \nu_m |t_-|) \exp(i \omega_- t_-) \, dt_-,
\end{equation}
where $\nu_m=\omega_m-\omega_{\mathrm{av.}}$ describes the resonant frequency difference contributed by the level $\mid m \rangle$ and the coefficients $C_m$ are complex numbers describing the transition dipoles associated with that level, 
\begin{equation}
C_m= \sigma_{TP} \langle \Phi_{\mu f} \mid m \rangle\langle m \mid \Phi_{g \mu} \rangle.
\end{equation}
It may be worth noting that the phases of these complex coefficients are an essential part of the time evolution previously expressed by $\hat{U}(|t_-|)$. The frequency representation makes it more difficult to identify the effects of time delays, but it allows for an easier identification of the contributions made by each intermediate level. 

The separate Fourier transforms in Eq.(\ref{eq:Gamma}) involve a sudden change of sign in the imaginary part of the function $\exp(-i \nu_m |t_-|)$. This change of sign is a consequence of the fact that there is no time evolution before the absorption of the first photon. Similar to the mathematics of linear response, this means that the Fourier transform results in resonant and off-resonant terms related to each other by Kramers-Kronig relations. The frequency components of the optimally absorbed state contributed by the level $\mid m \rangle$ are not limited to the resonant frequencies $\nu_m$ of that level. Instead, the result is given by a sum of resonant and off-resonant terms,
\begin{equation}
\Gamma(\omega_-) = \sum_m 2 C_m \left(\pi \delta(\omega_-+\nu_m)+ \pi \delta(\omega_--\nu_m) + \frac{i}{\omega_-+\nu_m} - \frac{i}{\omega_--\nu_m} \right).
\end{equation}
The off-resonant term can be summarized to emphasize the symmetry between positive and negative frequency differences,
\begin{equation}
\label {eq:KKR}
\Gamma(\omega_-) = \sum_m 2 C_m \left(\pi \delta(\omega_-+\nu_m)+ \pi \delta(\omega_--\nu_m) + i
\frac{2 \nu_m}{\nu_m^2 - \omega_-^2}\right).
\end{equation}
Eq.(\ref{eq:KKR}) shows that the resonant contributions of the intermediate levels occur at $\omega_-=\nu_m$. As expected, this corresponds to a two photon state with frequencies of $\omega_1=\omega_m-\omega_g$ and $\omega_2=\omega_f-\omega_m$. Since the Fourier transform does not impose a limit on the possible range of output frequencies, it is possible to obtain negative photon frequencies for $\omega_m>\omega_f$ or $\omega_m<\omega_g$. These results indicate that the infinite time resolution suggested by the time representation of the optimally absorbed two photon state is only possible because of the inclusion of unphysical states describing negative frequencies. As we will explain in the following, the oscillations derived in \cite{Sal98} as evidence for resonances at $\omega_m>\omega_f$ originated from this inclusion of unphysical frequencies and will not be observed in an actual experiment.

\section{Removal of unphysical negative frequency components}
\label{sec:negatives}


The Fourier transformation in Eq. (\ref{eq:FT}) is mathematically reversible if the complete range of frequencies $\omega_-=(\omega_2-\omega_1)/2$ between negative infinity and infinity is included. Since the frequency sum is given by $\omega_{gf}$, this frequency range includes negative frequencies for $|\omega_-|>\omega_{gf}/2$. It is very likely that time basis representations of entangled two photon states inadvertently include the effects of such negative frequency components. It is therefore important to exclude the effects of negative frequency components in theoretical descriptions of the TPA process. As mentioned above, this was not done in the case of \cite{Sal98} where the input state was described by a time domain wave function with a Fourier transform that included negative frequency components. The prediction that a time delay $\tau$ may result in oscillations at frequencies determined by levels above the final two photon excitation $\mid f \rangle$ was therefore based on the inclusion of negative energy photons that would de-exite the material upon their resonant absorption. Our analysis requires no assumptions about the two photon input state and allows us to separate the contributions from different intermediate levels in the frequency domain. It is therefore comparatively easy to exclude the unphysical negative frequencies in the contributions from each level. A corrected time representation of the optimally absorbed two photon state can then be derived by Fourier transforming the physically possible range of frequencies back into the time domain. As mentioned above, it is a straightforward matter to identify negative frequencies. For the frequency sum and the frequency difference of two photons, the individual frequencies are positive if $|\omega_1+\omega_2|>|\omega_1-\omega_2|$. Since the sum frequency of TPA is given by $\omega_+=\omega_1+\omega_2$ and $\omega_-$ is defined as one half of the frequency difference, we have to exclude all frequency differences with $|\omega_-|>\omega_{gf}/2$, leaving only those frequency differences in the interval $-\omega_{gf}/2<\omega_-<\omega_{gf}/2$.


If all intermediate levels lie in the interval of levels between ground state and final state, the resonant contributions are all physical and the temporal response will not be changed much from the one described in Sec. \ref{sec:dynamics}. As demonstrated in \cite{Svo19}, it is then possible to extract information about the electronic level structure of the absorbing medium from the effects of a time delay in the absorption of broad band entangled photon pairs. However, this may not be the case when the resonant contributions are unphysical. We therefore focus the following discussion on the case of virtual levels with energies larger than the two photon excited state, $\omega_m>\omega_f$($\nu_m>\omega_{gf}/2$). In this case the resonant contributions of the intermediate levels are all unphysical and need to be removed from the temporal profile of the optimally absorbed state. In practice, this means that we have to exclude the $\delta$ function contributions in Eq.(\ref{eq:KKR}), leaving only the off-resonant response terms,
\begin{equation}
\label {eq:off-resonant}
\Gamma(\omega_-) = \sum_m 4 i C_m \left(
\frac{\nu_m}{\nu_m^2 - \omega_-^2}\right).
\end{equation}
It should also be noted that this spectral wave function is only defined in the interval of physical frequencies, $-\omega_{gf}/2<\omega_-<\omega_{gf}/2$. 

The original dependence of the wave function on $t_-$ in Eq. (\ref{eq:timestate}) includes unphysical negative frequencies representing a time resolution that cannot be achieved in real physical systems. It is therefore necessary to correct the dependence of the absorption on the photon arrival time difference $t_-$ by removing the negative frequencies. The effect of this correction can be seen directly when the truncated expression for $\Gamma(\omega_-)$ is transformed back into the time representation. 
Since the Fourier transform is a linear operation it can be performed separately for each off-resonant contribution $m$. The final result is expressed by a sum of the individual contributions from each intermediate level $\mid m \rangle$,
\begin{equation}
\Gamma_\mathcal{FT}(t_-) = \sum_m 4 i C_m \gamma_m(t_-).
\end{equation}
Each contribution $\gamma_m(t_-)$ depends only on the resonant frequency of the intermediate level $\omega_m$ with respect to the two photon resonance. In terms of $\nu_m$ and $\omega_{gf}$, the analytical results of the Fourier transforms read
\begin{eqnarray}
\gamma_m(t_-)= &&\cos(\nu_m t_-)[\mbox{Ci}
((\nu_m+\omega_{gf}/2)t_-)-\mbox{Ci}((-\nu_m+\omega_{gf}/2)t_-)] \nonumber\\
&&-\sin(\nu_m t_-)[\mbox{Si}((\nu_m-\omega_{gf}/2)t_-)-\mbox{Si}((\nu_m+\omega_{gf}/2)t_-)],
\end{eqnarray}
where Si is the sine integral and Ci is the cosine integral. 
These are the corrected contributions of intermediate levels with $\omega_m>\omega_f$ to the arrival time difference dependence of the optimally absorbed wave function described by Eq.(\ref{eq:timestate}). Specifically, these contributions replace the original contributions of each intermediate level $\mid m \rangle$ given by the expansion of the unitary transformation shown in Eq.(\ref{eq:unitary}). These original contributions are described by complex oscillations proportional to $\exp(-i \nu_m |t_-|)$, representing the effects of resonant transitions between the levels. As expected, these resonant terms disappear completely after the negative frequency components have been removed. However, the real valued wave functions $\gamma_m(t_-)$ still depend on the precise value of the resonance $\nu_m$ of the intermediate level $\mid m \rangle$. It is therefore important to examine whether this dependence could be used to perform a kind of virtual level spectroscopy similar to the one proposed in \cite{Sal98}. 

\begin{figure}[th]
\begin{picture}(500,250)
\put(50,0){\makebox(400,180){
\scalebox{0.8}[0.8]{
\includegraphics{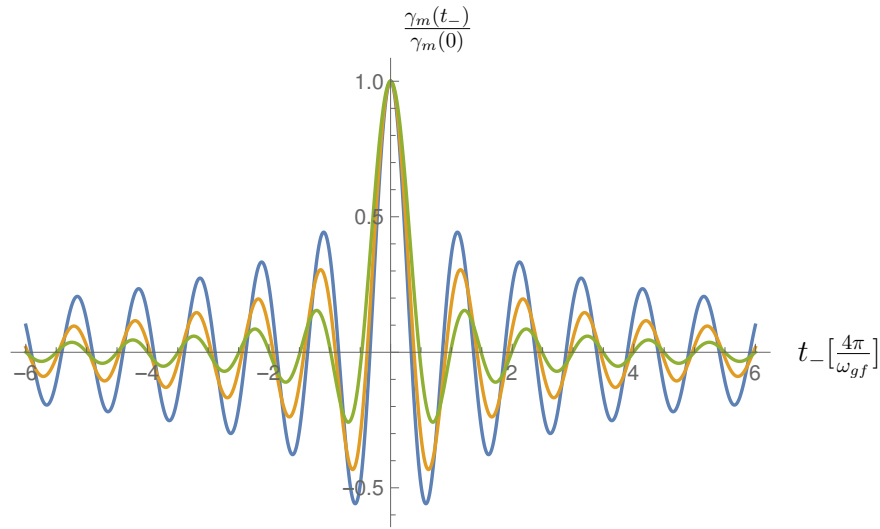}}}}
\end{picture}
\caption{\label{fig1}
Contribution $\gamma_m(t_-)$ to the time difference wave function of the optimally absorbed state for $\nu_m=1.02 \omega_{gf}/2$ (blue), $\nu_m=1.1 \omega_{gf}/2$ (orange) and $\nu_m=2 \omega_{gf}/2$ (green). For comparison, the wave function components are normalized to their value at $t_-=0$. The frequencies of the oscillations are close to $\omega_{gf}/2$ for all values of $\nu_m$. The only observable effect of $\nu_m$ is the increase in the amplitudes at $t_- > 2 \pi/\omega_{gf}$ for values of $\nu_m$ close to the minimal value of $\omega_{gf}/2$.
}
\end{figure}

Fig. \ref{fig1} shows the graphs of $\gamma_m(t_-)/\gamma_m(0)$ for different ratios of $\nu_m$ and $\omega_{gf}/2$. The graphs are normalized with respect to the amplitude at $t_-=0$ to compare their qualitative dependence on the time difference between the absorbed photons. For $\nu_m$ close to $\omega_{gf}/2$, the wave functions exhibit strong oscillations at $\omega_{gf}/2$ that continue over a long range of time differences. It may be tempting to identify these oscillations with the resonant frequency $\nu_m$, but the comparison with higher values of $\nu_m$ shows that the frequency of the oscillations changes only negligibly and remains at $\omega_{gf}/2$ for all values of $\nu_m>\omega_{gf}/2$. Fig. \ref{fig1} thus confirms that time delays $\tau$ cannot result in any oscillations of the TPA probability with frequencies of $\nu_m>\omega_{gf}/2$. The oscillations in the amplitudes of the optimally absorbed states associated with the intermediate level $\mid m \rangle$ originate from the abrupt cut-off of the spectrum at $\omega_-=\pm \omega_{gf}/2$. As $\nu_m$ increases, these oscillations drop to lower amplitudes compared to $\gamma_m(0)$ as the spectral wave functions approach a rectangular shape. 

\begin{figure}[th]
\begin{picture}(500,480)
\put(0,0){\makebox(500,480){
\scalebox{0.8}[0.8]{
\includegraphics{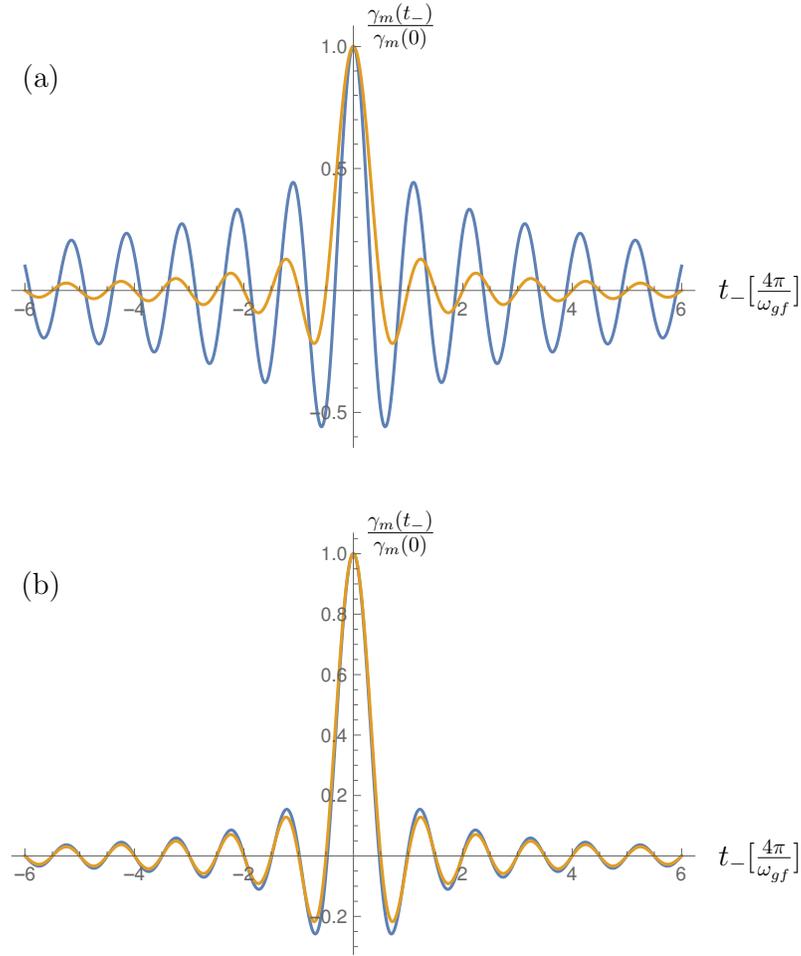}}}}
\end{picture}
\caption{\label{fig2}
Comparison of the exact analytical form (blue) of $\gamma_m(t_-)$ with its approximation (orange) using a sinc function for (a) $\nu_m=1.02 \omega_{gf}/2$ and (b) $\nu_m=2 \omega_{gf}/2$. The sinc function is a good approximation for all values of $\nu_m>2 \omega_{gf}/2$.
}
\end{figure}


For sufficiently high values of $\nu_m$, it is possible to approximate the spectral dependence by a rectangular function with the same area as the precise expression. This approximation corresponds to a sinc function in $t_-$ with the correct value at $t_-=0$. Specifically, the approximation reads
\begin{equation}
\gamma_m(t_-) \approx \log \left(\frac{\nu_m+\omega_{gf}/2}{\nu_m-\omega_{gf}/2}\right)
\frac{\sin(\omega_{gf} t_-/2)}{\omega_{gf} t_-/2}.
\end{equation}
Fig. \ref{fig2} shows a comparison of this approximation with the exact analytical result. For $\nu_m=\omega_{gf}$ ($\omega_m-\omega_f = \omega_{gf}/2$) the difference is rather small. We can therefore represent the effect of any number of intermediate levels with frequencies $\omega_m > \omega_f+\omega_{gf}/2$ by a simple summation of the coefficients, 
\begin{equation}
\Gamma_\mathcal{FT}(t_-) \approx \left(\sum_m 4 i C_m \log \left(\frac{\nu_m+\omega_{gf}/2}{\nu_m-\omega_{gf}/2}\right)\right) \; \frac{\sin(\omega_{gf} t_-/2)}{\omega_{gf} t_-/2}.
\end{equation}
The frequencies $\nu_m$ have no effect on the dependence on time difference $t_-$. Effectively, the sensitivity to time delays is determined by the TPA frequency $\omega_{gf}$, such that the TPA vanishes for time delays larger than $1/\omega_{gf}$. The optimally absorbed state for all off-resonant virtual transitions can be represented by a rectangular spectrum of all physical frequencies, $-\omega_{gf}/2<\omega_-<\omega_{gf}/2$. In time representation, this corresponds to a sinc function with a width of $4\pi/\omega_{gf}$, equal to the period of the average single photon frequency $\omega_{gf}/2$. Off-resonant virtual transitions are therefore ideal for the evaluation of broad band entangled states since the optimally absorbed two photon state represents the ideal limit of single cycle broad band entanglement.  

Our result shows that the signature of off-resonant virtual levels predicted by Saleh et al. \cite{Sal98} was an artefact of the time representation and does not occur in actual virtual state spectroscopy. Even in the case of frequencies $\omega_m$ close to the two photon excited level at $\omega_f$ the time difference dependence of the optimally absorbed state only reflects the cut off at negative frequencies. For most practical purposes it seems sufficient to think of the TPA process as essentially instantaneous. The time scale set by the TPA frequency ensures that the optimally absorbed state corresponds to a single cycle time difference wave function, the maximal amount of broad band entanglement that can be achieved in time-energy entanglement of photons.

\section{Conclusions}
\label{sec:conclude}

We have explained the temporal signatures of TPA by separating the sum frequency resonance from the time difference dependence of the optimally absorbed two photon state. It is then possible to understand the optimal temporal correlations between the photons in terms of the real time dynamics of the intermediate states. When transforming the optimally absorbed state into the frequency difference representation, each intermediate level contributes resonant and off-resonant terms related to each other by Kramers-Kronigs relations. In most cases the resonant terms will dominate the absorption process, reflecting the specific dynamics of the intermediate levels. However, the unrealistic time resolution implied by the time different basis results in unphysical negative frequencies that need to be removed from the result. For two photon absorbers without any resonances in the frequency range between the ground state frequency $\omega_g$ and the two photon excited state frequency $\omega_f$, the removal of unphysical frequencies results in a time difference dependence of the optimally absorbed wave function defined by a single time scale of $2 \pi/\omega_{gf}$, corresponding to a single period of the TPA frequency $\omega_{gf}$. Contrary to the result reported in \cite{Sal98}, the frequencies of virtual levels above the final two photon excited state do not show up in the time dependence of the optimally absorbed two photon state. Instead, two photon absorbers with no intermediate levels below the final state optimally absorb broad band entangled states with equal amplitudes for all frequencies between zero and $\omega_{gf}$. 

As our results show, the separation between sum frequency and time difference makes it much easier to characterize the material properties of two photon absorbers. By relating the absorption dynamics directly to the quantum coherence of the energy-time entangled two photon state, the role of entanglement in the absorption process can be explained directly in terms of the spectrum of available intermediate states. These insights might open up the way towards a more efficient characterization of broad band entanglement using TPA effects.

\begin{acknowledgments}
We would like to thank F. Schlawin and H. Oka for helpful discussions. This work has been supported by JST-CREST (JPMJCR1674), Japan Science and Technology Agency.
\end{acknowledgments}

\end{document}